\documentclass[
prd,twocolumn,
tightenlines,nofootinbib,
showpacs]{revtex4}
\usepackage{amssymb,latexsym}
\usepackage{amsmath,amsbsy,bbm}
\usepackage{epsfig,bm}
\usepackage{graphicx,comment}
\usepackage{color}
\DeclareMathOperator{\tr}{tr}

\begin{document}
\def\a{{\alpha}}
\def\b{{\beta}}
\def\d{{\delta}}
\def\D{{\Delta}}
\def\e{{\varepsilon}}
\def\g{{\gamma}}
\def\G{{\Gamma}}
\def\k{{\kappa}}
\def\l{{\lambda}}
\def\L{{\Lambda}}
\def\m{{\mu}}
\def\n{{\nu}}
\def\o{{\omega}}
\def\O{{\Omega}}
\def\S{{\Sigma}}
\def\s{{\sigma}}
\def\th{{\theta}}
\newcommand{\mnod}{\stackrel{\circ}{M}}

\def\ol#1{{\overline{#1}}}

\def\Dslash{\ol D\hskip-0.65em /}
\def\Dslashe{D\hskip-0.65em /}
\def\Pslash{\ol P\hskip-0.65em /}
\def\lslash{l\hskip-0.35em /}
\def\Pslashe{P\hskip-0.65em /}

\def\Dtslash{\tilde{D} \hskip-0.65em /}

\def\CPT{{$\chi$PT}}
\def\QCPT{{Q$\chi$PT}}
\def\PQCPT{{PQ$\chi$PT}}
\def\tr{\text{tr}}
\def\str{\text{str}}
\def\diag{\text{diag}}
\def\order{{\mathcal O}}

\def\cC{{\mathcal C}}
\def\cB{{\mathcal B}}
\def\cT{{\mathcal T}}
\def\cQ{{\mathcal Q}}
\def\cL{{\mathcal L}}
\def\cO{{\mathcal O}}
\def\cA{{\mathcal A}}
\def\cQ{{\mathcal Q}}
\def\cR{{\mathcal R}}
\def\cG{{\mathcal G}}
\def\cH{{\mathcal H}}
\def\cF{{\mathcal F}}
\def\cW{{\mathcal W}}
\def\cM{{\mathcal M}}
\def\cD{{\mathcal D}}
\def\cN{{\mathcal N}}
\def\cP{{\mathcal P}}
\def\cK{{\mathcal K}}
\def\Qt{{\tilde{Q}}}
\def\Dt{{\tilde{D}}}
\def\St{{\tilde{\Sigma}}}
\def\cBt{{\tilde{\mathcal{B}}}}
\def\cDt{{\tilde{\mathcal{D}}}}
\def\cTt{{\tilde{\mathcal{T}}}}
\def\cMt{{\tilde{\mathcal{M}}}}
\def\At{{\tilde{A}}}
\def\cNt{{\tilde{\mathcal{N}}}}
\def\cOt{{\tilde{\mathcal{O}}}}
\def\cPt{{\tilde{\mathcal{P}}}}
\def\cI{{\mathcal{I}}}
\def\cJ{{\mathcal{J}}}
\def\cb{{\cal B}}
\def\cbb{{\overline{\cal B}}}
\def\ct{{\cal T}}
\def\ctt{{\overline{\cal T}}}

\def\eqref#1{{(\ref{#1})}}

\preprint{UMD-40762-442}
 
\title{Connected Parts of Decuplet Electromagnetic Properties}

\author{Brian~C.~Tiburzi}
\email[]{bctiburz@umd.edu}
\affiliation{Maryland Center for Fundamental Physics, Department of Physics, University of Maryland, College Park, MD 20742-4111, USA}

\date{\today}


\begin{abstract}
We determine the electromagnetic properties of decuplet resonances
using chiral perturbation theory at next-to-leading order. 
Utilizing a partially quenched charge matrix, 
we isolate and remove the quark disconnected contractions.
This allows us to compare the physical meson loop contributions to the connected
contributions, of which the latter are currently calculated using lattice QCD. 
Finally we determine linear combinations of 
decuplet resonance 
and hyperon electromagnetic properties that are exactly independent of disconnected 
contractions in the isospin limit. 
\end{abstract}

\maketitle

\section{Introduction}

Lattice QCD simulations continue to make dramatic progress
towards addressing quantitatively the physics of strong 
interactions in the non-perturbative regime, 
see~\cite{DeGrand:2006aa} for a comprehensive overview. 
In order to address the chiral extrapolation of observables
determined from partially quenched lattice QCD simulations
(which employ differing values for  valence and sea quark masses), 
partially quenched chiral perturbation theory (\CPT)
has been developed~\cite{Bernard:1993sv,Sharpe:2000bc,Sharpe:2001fh}. 
These techniques are applicable in a wider context, 
namely any situation for which valence and sea quark contributions must be distinguished. 
This is the case, 
for example, 
in mixed action simulations which employ different fermion 
discretizations for valence and sea quarks. 
Despite tuning valence and sea quark masses to be 
identical, masses of mixed mesons made of one valence and one sea quark
are not protected from additive renormalization, 
and partially quenched \CPT\ calculations are needed to control systematics, 
see, e.g.~\cite{Bar:2002nr,Bar:2005tu,Tiburzi:2005is}.

An example relevant to the present work is that of the lattice 
determination of connected parts of correlation functions. 
Operator self-contractions are notorious for their statistical noise, 
and are often omitted from lattice calculations. 
The systematic effects of this approximation can be addressed using 
partially quenched \CPT. 
We will focus our consideration on the electromagnetic properties
of the delta resonance. 
Connected parts of other observables in some cases
have been explicitly computed%
~\cite{Tiburzi:2004mv,Detmold:2005pt,Tiburzi:2005is,Chen:2006gg,Jiang:2008ja}, 
or can easily be derived from existing partially quenched results%
~\cite{Chen:2001yi,Beane:2002vq,Arndt:2003ww,Arndt:2003we,Arndt:2003vd}. 
The electromagnetic properties of delta resonances 
can be accessed indirectly using electromagnetic excitation 
of  nucleons, see%
~\cite{Nefkens:1977eb,Bosshard:1991zp,LopezCastro:2000ep,Kotulla:2002cg,Pascalutsa:2004je,Pascalutsa:2007wb}.
With recent experimental work at MAMI~\cite{Kotulla:2008zz}, there has been renewed theoretical interest. 
Model independent results for decuplet electromagnetic properties
have been deduced using \CPT%
~\cite{Butler:1993ej,Banerjee:1995wz,Pascalutsa:2004je,Pascalutsa:2007wb}.
Beyond investigations using the quenched approximation~\cite{Leinweber:1992hy,Lee:2005ds},
dynamical lattice QCD computations have been recently carried out,
but with the exclusion of quark disconnected contractions%
~\cite{Alexandrou:2008bn,Aubin:2008qp,Alexandrou:2009hs}.

In this note, we revisit the computation of decuplet resonance electromagnetic properties using \CPT. 
Our focus is on the connected part of such properties that have been investigated recently using lattice QCD simulations. 
It is not widely appreciated that partially quenched \CPT\
can be utilized to address the omitted current self-contractions in these lattice calculations, 
e.g.%
~\cite{Alexandrou:2008bn} omits discussion of such systematic errors, 
while%
~\cite{Aubin:2008qp} suggests that perhaps one should compare lattice results to the quark model. 
We determine the connected contributions by turning off the electric charges of the sea quarks. 
The observables we consider are the charge radii, 
magnetic moments, 
and electric quadrupole moments. 
The decuplet magnetic octupole moments vanish at next-to-leading order because there are neither local nor one-loop 
$d$-wave couplings to the photon~\cite{Butler:1993ej}. 
This fact is not altered by quenching or partial quenching~\cite{Arndt:2003we}.
We also correct the absolute normalization of the magnetic moments used in~\cite{Butler:1993ej,Jenkins:1992pi},
and the erroneous loop integrals appearing in~\cite{Arndt:2003we}.

Our main observation is that because partially quenched \CPT\ 
tracks the flow of sea quarks, we can use this theory to determine
the chiral behavior of the quark connected parts of electromagnetic
current matrix elements. 
Consider the 
$SU(6|3)$ 
fundamental vector of quarks
\begin{equation}
\psi^T
= 
(u, d, s, j, l, r, \tilde{u}, \tilde{d}, \tilde{s})
,\end{equation}
where in addition to the three light quark flavors, 
$(u,d,s)$, 
there appear three ghost quarks 
$(\tilde{u}, \tilde{d}, \tilde{s})$. 
These ghosts are spin-$1/2$ quarks of the wrong statistics
so that when made degenerate with 
$(u,d,s)$ 
respectively the closed light quark loops vanish. 
Thus using  
$(u,d,s)$ 
in hadronic interpolating fields renders them truly valence quarks. 
With such valence hadrons, the remaining quarks 
$(j,l,r)$
only appear in disconnected loops and are hence sea quarks. 
While in general the extension of electric charges to partially 
quenched theories is not unique%
~\cite{Golterman:2001qj,Chen:2001yi}, 
there is an obvious choice corresponding to the actual lattice computations%
~\cite{Tiburzi:2004mv,Detmold:2005pt} 
\begin{equation} \label{eq:charge}
\cQ = \diag ( q_u, q_d, q_s, q_j, q_l, q_r, q_u, q_d, q_s)
.\end{equation}
Here the ghost quarks are given identical charges to their valence counterparts,
so that the only disconnected operator insertions arise from couplings to the sea  quarks. 
The omission of quark disconnected contributions then corresponds to vanishing sea quark charges, 
$q_j = q_l = q_s = 0$.

Meson loop diagrams make important contributions to the electromagnetic properties of hadrons. 
Imagine we have a 
$(j\ol{d)}$-valence-sea 
meson in a one-loop diagram for some electromagnetic observable. 
The 
$j$-quark
is the sea counterpart to the 
$u$-quark and so 
ordinarily the photon coupling to this meson is set by the charge of the 
$\pi^+$, 
namely 
$Q = 1$.
With disconnected contractions omitted, 
$q_j = 0$, 
and the charge of the 
$(j\ol{d})$-meson becomes
$Q = q_j - q_d = 1/3$. 
The charges of some loop mesons are reduced by omitting disconnected diagrams. 
Na\"ively the contributions from charged meson loops are decreased in magnitude. 
Surprisingly quite the opposite can also be true: 
meson loop diagrams can be enhanced by neglecting disconnected operator
contractions. 
This is because while some loop meson charges are decreased, 
others are suddenly turned on.   
In the partially quenched theory there is now also a
$(j \ol u)$-meson, 
which in \CPT\ is electrically neutral (it is part of a propagating $\pi^0$).
Dropping disconnected contributions, 
however,  
the charge of this meson no longer vanishes,
$Q = q_j - q_u = - 2/3$,
and the photon couples to it through loop diagrams.
Of course, 
partially quenched \CPT\ must be utilized to determine
the group theoretic and charge factors for the various loop meson 
contributions.

At this piont, 
we employ Eq.~\eqref{eq:charge} with vanishing sea quark charges 
and revisit the calculation in~\cite{Arndt:2003we}.
This enables us to determine the connected parts of decuplet
electromagnetic properties. 
We refer to this computation as connected \CPT, 
leaving \CPT\ to refer to the computation using Eq.~\eqref{eq:charge} with the physical values of the quark charges, 
i.e.~$q_u = q_j = 2/3$, $q_d = q_s = q_l = q_r = -1/3$.  
Because the setup for the computation parallels that 
of~\cite{Arndt:2003we}, we refer to that work for specific details. 
We present the connected parts of the decuplet charge radii, 
magnetic moments, 
and electric quadrupole moments. 
In doing so, 
moreover, 
we correct the erroneous loop integrals appearing in~\cite{Arndt:2003we}.

The electric charge radius is defined in terms of the slope of the 
charge form factor at zero momentum transfer, 
\begin{equation}
< r_E^2>
= 
6 \frac{d^2 G_{E0}(q^2)}{dq^2}  \Bigg|_{q^2 = 0}
.\end{equation}
For the decuplet state $T$, e.g.~$T = \D^{++}$, 
the charge radius takes the form
\begin{eqnarray}
< r_E^2>^T 
&=&
\frac{6 Q^T c_c}{(4\pi f)^2}
-
\frac{1}{27} \frac{81 + 25 \cH^2}{(4 \pi f)^2}
\sum_{\phi = \pi, K} A_\phi^T \, 
\log \frac{m_\phi^2}{\mu^2}
\notag \\
&& \phantom{sp}
-
\frac{5}{3} \frac{\cC^2}{(4 \pi f)^2}
\sum_{\phi = \pi, K} A_\phi^T \, 
\cG( - \D, m_\phi, \mu)
\label{eq:radii}
,\end{eqnarray}
where $Q^T$ is the charge of the decuplet state $T$, 
e.g.~for the $\D^{++}$, we have $Q^{\D^{++}} = 2$, 
$c_c$ is the unknown local contribution, 
$\D$ is the mass splitting between the decuplet and octet baryons, 
and the non-analytic function $\cG(\d, m,\mu)$ is 
defined in~\cite{Butler:1993ej}. 
In our conventions, 
the pion decay constant is given by 
$f = 132 \, \texttt{MeV}$. 
Appearing in the expression for charge radii 
are coefficients $A_\phi^T$ for contributing loop 
mesons. These have been tabulated in Table~\ref{t:A}. 
The tabulated values show the effect of neglecting disconnected 
current insertions. 
For example, the pion loop contributions to the 
$\D^+$
are doubled in the connected case, 
while those for the 
$\D^0$ 
are eliminated.
On the other hand, 
\CPT\ loop contributions for the 
$\S^{*, 0}$ 
should vanish but are turned on in connected \CPT.

\begin{table}
\caption{The loop coefficients $A_\phi^T$ for decuplet states in \CPT.
Listed are \CPT\ and connected \CPT\ values.}
\begin{ruledtabular}
\begin{tabular}{l | c c | c  c  }
$T$
& \multicolumn{2}{c |}{\CPT\ $\quad$}  &  \multicolumn{2}{c}{ connected~\CPT\ $\quad$} \\
$\qquad \quad \phi$
& $\pi$ & $K$  &   $\pi$   &   $K$  \\
\hline 
$\Delta^{++}$       
& $1$ & $1$  
& $\frac{4}{3}$ &  $\frac{2}{3}$ 
\\
$\Delta^{+}$        
& $\frac{1}{3}$ & $\frac{2}{3}$ 
& $\frac{2}{3}$ & $\frac{1}{3}$ 
\\
$\Delta^{0}$        
& $-\frac{1}{3}\phantom{-}$ & $\frac{1}{3}$  
& $0$ & $0$ 
\\
$\Delta^{-}$        
& $-1\phantom{-}$ & $0$ 
& $-\frac{2}{3}\phantom{-}$ & $-\frac{1}{3}\phantom{-}$ 
\\
$\Sigma^{*,+}$       
& $\frac{2}{3}$ & $\frac{1}{3}$  
& $\frac{8}{9}$ & $\frac{2}{9}$ 
\\
$\Sigma^{*,0}$       
& $0$ & $0$  
& $\frac{2}{9}$ & $-\frac{1}{9}\phantom{-}$ 
\\	
$\Sigma^{*,-}$      
& $-\frac{2}{3}\phantom{-}$ & $-\frac{1}{3}\phantom{-}$  
& $- \frac{4}{9}\phantom{-}$ & $ - \frac{4}{9}\phantom{-}$ 
\\
$\Xi^{*,0}$         
& $\frac{1}{3}$ & $-\frac{1}{3}\phantom{-}$  
& $\frac{4}{9}$ & $-\frac{2}{9}\phantom{-}$ 
\\
$\Xi^{*,-}$          
& $-\frac{1}{3}\phantom{-}$ & $-\frac{2}{3}\phantom{-}$  
& $-\frac{2}{9}\phantom{-}$ & $-\frac{5}{9}\phantom{-}$   
\\
$\Omega^-$          
& $0$ & $-1\phantom{-}$   
& $0$ & $-\frac{2}{3}\phantom{-}$ 
\\
\end{tabular}
\end{ruledtabular}
\label{t:A}
\end{table} 

To define magnetic moments of decuplet resonances, 
we use the standard identification,
namely the moment is defined by the 
$z$-component
of the magnetic moment operator,
$\vec{M}$, 
in the state of maximal 
$z$-angular momentum
\begin{equation} \label{eq:defn}
\mu^{j = \frac{3}{2}} 
\equiv 
\Big\langle j = \frac{3}{2}, m = \frac{3}{2} \Big| M_z \Big| j = \frac{3}{2}, m = \frac{3}{2} \Big\rangle
.\end{equation}
The leading tree-level \CPT\ operator is contained in the Lagrangian~\cite{Jenkins:1992pi}
\begin{equation} \label{eq:MM}
\cL
=
i \frac{e \mu_c}{M_N} Q^T \, \ol T {}^\mu T^\nu F_{\mu \nu}
,\end{equation}
with $F_{\mu \nu} = \partial_\mu A_\nu - \partial_\nu A_\mu$, 
and $M_N$ as the nucleon mass. 
Using the explicit form of the normalized Rarita-Schwinger spinor in Eq.~\eqref{eq:defn},
we find Eq.~\eqref{eq:MM} leads to magnetic moments of size:
$2 Q^T \mu_c$, 
in units of nuclear magnetons, 
$\mu_N =  e / 2 M_N $.
Using the multipole decomposition of the 
spin-$3/2$ current matrix elements~\cite{Nozawa:1990gt}, 
one confirms this result~\cite{Arndt:2003we}.

In units of nuclear magnetons, 
the magnetic moment  of the decuplet state 
$T$ 
is given by the expression
\begin{eqnarray}
\mu^T
&=& 
2 \mu_c Q^T
+
\frac{4 M_N \cH^2}{9 (4 \pi f)^2}
\sum_{\phi = \pi, K}
A_\phi^T \,
\pi \, m_\phi
\notag \\
&& \phantom{spac}
+
\frac{2 M_N \cC^2}{(4 \pi f)^2}
\sum_{\phi = \pi, K}
A_\phi^T \,
\cF(-\D, m_\phi, \mu)
\label{eq:moments}
.\end{eqnarray}
The coefficients for loop mesons $A_\phi^T$
are identical to those in Eq.~\eqref{eq:radii}, 
while the non-analytic function
$\cF(\d,m,\mu)$
is given in~\cite{Butler:1993ej}. 
The same comparison of \CPT\ results to connected
\CPT\ results can thus be made.

Finally
for the electric quadrupole moments, 
we determine the quadrupole form factor at zero
momentum transfer. 
The quadrupole moment of the decuplet state 
$T$
is given by
\begin{eqnarray}
G_{E2}(0) 
&=&
Q^{T} Q_c \frac{M_T^2}{( 4 \pi f)^2}
- 
\frac{8}{27} \frac{M_T^2 \cH^2}{( 4 \pi f)^2}
\sum_{\phi = \pi, K} A_\phi^T \, \log \frac{m_\phi^2}{\mu^2}
\notag \\
&& \phantom{spac}
+ 
\frac{2}{3} \frac{M_T^2 \cC^2}{(4 \pi f)^2}
\sum_{\phi = \pi, K} A_\phi^T \,  \cG ( - \D, m_\phi, \mu)
,\end{eqnarray}
where $Q_c$ is the unknown local contribution.

When one studies the connected part of nucleon current matrix elements using lattice QCD, 
one can form isovector combinations of electromagnetic observables
that are exactly independent of sea quark charges in the limit of strong isospin. 
There are analogous quantities for decuplets (and hyperons) that do not require
disconnected contributions to make physical predictions. 
To exploit isospin symmetry, 
we package the octet and decuplet states into representations of $SU(2)_V$%
~\cite{Beane:2003yx,Tiburzi:2008bk}. 
Consider two baryons (of the same spin) that are elements of 
the same isospin multiplet, 
e.g.~$\D^{++}$ and $\D^{+}$ are both in the quartet of deltas, 
or $\Sigma^+$ and $\Sigma^-$ are both in the triplet of spin-$1/2$ sigmas. 
Label these states by their isospin $j$, and $z$-components $m$ and $m'$. 
Now consider the difference of their current matrix elements
\begin{equation} \label{eq:d}
\d (m, m')
\equiv
\langle j , m | J_\mu | j , m \rangle
- 
\langle j, m' | J_\mu | j, m' \rangle
.\end{equation}
The electromagnetic current can be written in terms of isoscalar and isovector contributions, 
\begin{equation}
J_\mu = T^0_\mu + T^{1,0}_\mu
,\end{equation}
where
$T^0_\mu = \frac{1}{6} \ol u \gamma_\mu u + \frac{1}{6} \ol d \gamma_\mu d - \frac{1}{3} \ol s \gamma_\mu s$, 
and 
$T^{1,0}_\mu = \frac{1}{2} \ol u \gamma_\mu u - \frac{1}{2} \ol d \gamma_\mu d$. 
As a consequence of the Wigner-Eckart theorem, 
the isoscalar contribution must drop out
of $\d (m, m')$. 
While the isovector part of the current does not cancel out of the difference, each matrix element of 
$T^{1,0}_\mu$
does not have disconnected operator contractions if we take the strong isospin limit. 
Thus all differences of the form $\d (m, m')$ in Eq.~\eqref{eq:d} are exactly independent of sea quark charges provided $m_u = m_d$. 
Values for $A_\phi^T$ coefficients in Table~\ref{t:A} corroborate this fact at one-loop order in the chiral expansion.

The matrix element differences just discussed are exactly independent
of disconnected electromagnetic current contributions. 
There is, 
however,  
additional physical information that can be deduced from 
connected lattice QCD correlation functions by using the 
$SU(3)$ 
chiral expansion. 
We again group the hyperons into isospin multiplets. 
Consider the differences in particle electromagnetic properties between two different isospin multiplets.
The extracted differences in magnetic moments, 
for example, 
will not be directly comparable to experiment.
Despite this fact, 
the first low-energy constants omitted by not calculating 
disconnected contractions are at fourth order in the 
$SU(3)$ 
chiral expansion.
The reason for this stems from the traceless condition on the electric charge matrix,
as we now explain.

Operators in \CPT\ involving the trace of the charge matrix 
(or more correctly the supertrace in partially quenched theories) 
account for local interactions of the photon with the sea quarks. 
Such operators are turned off when only connected diagrams are calculated on the lattice. 
Because 
$\str ( \cQ ) = 0$, 
there are no leading-order operators depending on the sea quark charges, 
hence no local terms are omitted by calculating only the connected parts. 
At second order, 
there are operators depending on the sea quark charges that 
are proportional to
$\str ( m_q \cQ )$. 
A quark mass insertion costs two powers of the small expansion
parameter in \CPT\ power counting. 
This coupling to the sea quarks, however, is identical for each member of the
$SU(3)$
multiplet, 
hence differences between electromagnetic properties are free of 
disconnected local contributions at second order in the chiral expansion.

To find quark disconnected local terms contributing to differences 
of electromagnetic properties, we must go to fourth order. 
For example consider the following operator 
\begin{equation} \label{eq:O}
\cO = ( \ol T {}^\mu \, m_q \, T^\nu ) \, \str ( m_q \cQ ) F_{\mu \nu}
,\end{equation}
for the case of the decuplet baryons, with similar operators for the octet baryons.
When multiplied by an unknown low-energy constant, 
this operator makes contributions to decuplet magnetic moments. 
The action of this operator is different between isospin multiplets,
and consequently contributions that stem from it 
will not cancel in magnetic moment differences.
The operator in Eq.~\eqref{eq:O}, 
however, 
is turned off when one calculates only connected contributions.
Because the operator is at fourth order, 
the error introduced by omitting disconnected diagrams 
in the difference of magnetic moments
should be small provided the $SU(3)$ expansion is under control.

The differences in electromagnetic properties of particles from different isospin multiplets 
(but identical spins) 
are thus sensitive to all physical low-energy constants up to terms that 
scale generically as 
$m_\eta^4 / (2 \sqrt{2} \pi f)^4 \sim 0.06$. 
This fact is particularly useful for studying the connected parts of
$\mu^{\D^{++}} - \mu^{\O^-}$, 
and 
$\mu^{\S^+} - \mu^{\L \S}$
in lattice QCD, 
where 
$\mu^{\L \S}$ 
is the magnetic dipole transition moment between the 
$\L$ 
and 
$\S^0$ 
baryons.
To study such differences using lattice data for the connected contributions, 
one can determine the unknown low-energy constants,  
and input these values into \CPT\ expressions to make physical predictions.

Lastly we discuss an oddity of the decuplet electromagnetic properties relating to quenching. 
This is particularly relevant given claims from a recent quenched lattice QCD study~\cite{Boinepalli:2009sq}.
The loop mesons contributing to the decuplet current matrix elements are entirely of the valence-sea variety. 
If dynamical quarks are quenched, so too are the loop contributions calculated here. 
In the quenched case, 
there are unphysical contributions from quenched hairpins~\cite{Savage:2001dy}. 
These have been calculated explicitly, 
and shown to be the only contributions at next-to-leading order in the 
quenched theory of decuplet electromagnetic properties~\cite{Arndt:2003we}. %
To use these quenched \CPT\ results, 
one must first correct the erroneous loop integrals in~\cite{Arndt:2003we}. 
The following replacement to Eq.~(41) of that work corrects the quenched results
\begin{equation}
I(m_1,m_2, \D_1, \D_2, \mu) \to
I(m_1,m_2,\D_1 - \D, \D_2 - \D, \mu)
.\end{equation}
While it is true that decuplet masses retain parts of \CPT\ pion loop contributions
despite quenching~\cite{Tiburzi:2004rh,Tiburzi:2005na} 
(as happens for the nucleon~\cite{Labrenz:1996jy}), 
none of these contributions remain for 
their electromagnetic properties. 
Instead they only inherit artifacts of quenching. 
Fortunately lattice QCD has progressed beyond the quenched approximation.

\begin{acknowledgments}
We thank Paulo F.~Bedaque for discussion. 
This work is supported by the 
U.S.~Department of Energy, Grant 
No.~DE-FG02-93ER40762. 
\end{acknowledgments}

\bibliography{hb}

\end{document}